# CPIA Dataset: A Comprehensive Pathological Image Analysis Dataset for Self-supervised Learning Pre-training

Nan Ying, Yanli Lei, Tianyi Zhang, Shangqing Lyu, Chunhui Li, Sicheng Chen, Zeyu Liu, Yu Zhao, and Guanglei Zhang, *Member, IEEE*

***Abstract*—Pathological image analysis is a crucial field in computer-aided diagnosis, where deep learning is widely applied. Transfer learning using pre-trained models initialized on natural images has effectively improved the downstream pathological performance. However, the lack of sophisticated domain-specific pathological initialization hinders their potential. Self-supervised learning (SSL) enables pre-training without sample-level labels, which has great potential to overcome the challenge of expensive annotations. Thus, studies focusing on pathological SSL pre-training call for a comprehensive and standardized dataset, similar to the ImageNet in computer vision. This paper presents the comprehensive pathological image analysis (CPIA) dataset, a large-scale SSL pre-training dataset combining 103 open-source datasets with extensive standardization. The CPIA dataset contains 21,427,877 standardized images, covering over 48 organs/tissues and about 100 kinds of diseases, which includes two main data types: whole slide images (WSIs) and characteristic regions of interest (ROIs). A four-scale WSI standardization process is proposed based on the uniform resolution in microns per pixel (MPP), while the ROIs are divided into three scales artificially. This multi-scale dataset is built with the diagnosis habits under the supervision of experienced senior pathologists. The CPIA dataset facilitates a comprehensive pathological understanding and enables pattern discovery explorations. Additionally, to launch the CPIA dataset, several state-of-the-art (SOTA) baselines of SSL pre-training and downstream evaluation are specially conducted. The CPIA dataset along with baselines is available at https://github.com/zhanglab2021/CPIA_Dataset.***

## I. INTRODUCTION

CANCER has emerged as a significant threat to human health, with approximately 1,958,310 new cases and 609,820 fatalities anticipated in 2023 according to the American Cancer Society [1]. In clinical practice, pathological examination serves as the gold standard for diagnoses. Moreover, clinical records and physical examination alone may be inadequate for accurate diagnosis of certain non-communicable diseases [2]. Therefore, pathological examination informs treatment guidelines and prognosis assessment [2]. A 2016 interview-based study in the USA and Germany revealed that 66% of clinical decisions were dependent on the results of pathology and laboratory medicine [3].

However, exclusively depending on human evaluation retains limitations and dilemmas. Firstly, these diagnoses are primarily based on pattern recognition within images by pathologists and the interpretation of such patterns in a wider range of patients [4]. Therefore, the reproducibility of such interpretations among different pathologists is unsatisfactory [4]. Secondly, errors and misdiagnoses are unavoidable in pathological examinations [5]. Finally, an increasing shortage of pathologists persists in numerous countries, leading to overburdened workloads and reduced efficiency. Such situations are especially severe in developing regions [6]. Thus, life-saving diagnosis improvement needs more automatic and accessible solutions.

In recent years, artificial intelligence (AI) has revolutionized various fields, such as computer-aided diagnosis in pathological image analysis aiming to address the aforementioned challenges effectively [7][8]. Deep learning has been widely applied to pathological tasks such as distinguishing benign and

This work was partially supported by the National Natural Science Foundation of China (No. 62271023), the Beijing Natural Science Foundation (No. 7202102), and the Fundamental Research Funds for Central Universities.

N. Ying, Y. Lei and T. Zhang contributed equally to this work. Corresponding Author: Guanglei Zhang (e-mail: guangleizhang@buaa.edu.cn).

N. Ying, Y. Lei, T. Zhang, Z. Liu, and G. Zhang are with the Beijing Advanced Innovation Center for Biomedical Engineering, School of Biological Science and Medical Engineering, Beihang University, Beijing 100191, China (e-mails: {20101014, 20373207, zhangtianyi, zeyuliu, guangleizhang}@buaa.edu.cn).

S. Lyu is with the Bioinformatics Institute (BII), Agency for Science, Technology and Research (A*STAR), Singapore 138671, Singapore (e-mail: shangqinglyu@outlook.com).

C. Li is with the Department of Artificial Intelligence, Nanjing University, Nanjing 210093, China (e-mail: lich@smail.nju.edu.cn).

S. Chen is with the School of Microelectronics, Xi'an Jiaotong University, Xi'an 710049, China (e-mail: wubanbao@163.com).

Y. Zhao is with the Department of Pathology, Peking Union Medical College Hospital, Beijing 100006, China (e-mail: rain986532@126.com).

malignant tumors, cancer grading, nuclear and glandular segmentation, and mitotic evaluation [9-13].

The traditional deep learning models depend on massive labeled data for the aforementioned tasks. However, the scarcity of such labels in the pathological field lefts a significant hindrance to their effectiveness [8]. By initializing models with pre-trained weights, transfer learning can provide pre-knowledge to alleviate such dilemmas.

Pre-training can improve model recognition and downstream task performance [14][15]. In this process, the pre-training datasets play a crucial role. For example, natural image analysis has experienced unprecedented growth with the support of large-scale public pre-training datasets like ImageNet and COCO [16][17]. Such datasets for pre-training are lacking in the field of pathological image analysis, forcing current researchers to initialize models based on various natural images. In the pre-training stage, the model learns a general feature representation from natural images, which is then utilized for downstream pathological image analysis tasks. Although improved results have been achieved to a certain extent, the huge gap between natural and pathological images has become a stumbling block to further researches [18-20]. Therefore, a comprehensive pre-training dataset based on pathological images is needed.

The traditional supervised pre-training relies on large amounts of labeled data. However, the annotations of pathological images are expensive, which require significant workloads of experts to analyze pathological images. The scarcity of labeled pathological data conflicts with the need of traditional supervised pre-training. As a trending pre-training technique, self-supervised learning (SSL) is a form of unsupervised learning where the data provides the supervision, unsealing the field from limited annotations [21]. Specifically, a pretext task is defined using unlabeled pre-training images to train models before supervised downstream task training, thereby improving model robustness and reducing data bias disturbance [22]. After learning the general features of upstream data, the models are transferred to particular downstream tasks. Beyond improvements observed in downstream tasks, the SSL pre-training process also aligns with the supervision-lack characteristic of pathological images [14][15].

Although SSL enables label-free general encoding, the pre-training process highly relies on the comprehensive and standardized datasets. However, there are currently two major challenges in the construction of SSL pathological image analysis datasets. Firstly, the existing datasets are difficult to meet the requirements of comprehensiveness and diversity. Secondly, the large-scale pathological pre-training dataset demands a standardization processing workflow, while the diverse features and complex sampling conditions among the samples lead to great challenges.

Regarding the first challenge, most existing datasets are limited to specific diseases with small volumes, which is insufficient to support general pathological knowledge learning [8]. For example, Camelyon17 is a dataset for detecting and classifying breast cancer metastasis in lymph nodes with 1,000 images, and SICAPv2 is a prostate histology dataset with both annotations of global Gleason scores and patch-level Gleason grades with 18,783 images [23][24]. Thus, a comprehensive pathological SSL pre-training dataset is expected, similar to the ImageNet dataset of natural images, which includes approximately 14,000,000 images [16]. Furthermore, to lay a solid foundation and advance the development of pathological image analysis, the dataset needs to cover a wide range of diseases and organs/tissues [19]. Additionally, the larger the SSL pre-training data magnitude is, the more significant the model performance would improve.

Accordingly, this paper proposes a comprehensive pathological image analysis (CPIA) dataset for SSL pre-training to achieve the comprehensive and large-scale requirements. The CPIA dataset comprises 21,427,877 images, encompassing over 48 organs/tissues and approximately 100 disease types. It primarily includes two pathological sample types: whole slide images (WSIs) without annotation and characteristic regions of interest (ROIs) that have been aligned after clearing, standardization, and task-clustering. To the best of our knowledge, the comprehensiveness and enormity of the CPIA dataset surpass all existing datasets for pathological image analysis. The wide-covering histopathology and cytopathology data enable transfer learning to acquire comprehensive pathological knowledge, rather than being restricted to a particular organ or disease. Therefore, the CPIA dataset solves the first challenge in the construction of SSL pathological image analysis datasets.

Regarding the second challenge, this paper proposes a workflow of strict aligning and systematic composition. There are 103 published pathological datasets used in constructing the CPIA dataset. And these dataset variations in task types, image quality, and construction frameworks lead to great challenges. To solve these challenges, we first transform WSIs to the same scale with a uniform area of tissue slides based on a unified micron per pixel (MPP). Moreover, we combine the diagnosis habits of pathologists with dataset construction by designing subsets of different scales. Lastly, the different ROI public datasets are grouped with the corresponding scale subsets with the supervision of experienced senior pathologists.

In previous studies, numerous datasets are limited to a single predefined scale tailored for a specific task [25][26]. Neglecting this diverse information available at different scales leads to information loss, while the CPIA dataset quantitatively incorporates multi-scale images by building different scale subsets [27][28]. In fact, pathological images at different scales contain varying information, collectively determining sample properties and individually corresponding to different downstream task types [8]. For the first time in SSL, we integrate the diagnosis habits of pathologists into dataset construction, producing a dataset with better clinical relevance.

Additionally, the comprehensive coverage of the CPIA dataset facilitates the full and objective evaluations of model performance. For the efficiency and convenience of evaluation, this paper also introduces the CPIA-Mini dataset for SSL pre-training, along with several independent downstream evaluation datasets. Specifically, the CPIA-Mini dataset is a lightweight and categorically balanced CPIA fraction, which includes 3,383,970 images. It is non-overlapping to the validation and test subsets of downstream evaluation datasets. This feature facilitates the exploration of the performance of various state-of-the-art (SOTA) SSL pre-training baselines. The baseline experiments are reported based on the CPIA-Mini dataset.

In summary, the contribution can be concluded into the following five aspects:

- Introduction of the CPIA dataset for SSL pre-training, characterized by its large-scale and wide-covering, including 21,427,877 images, over 48 organs/tissues, and approximately 100 disease types.
- Systematic data processing and strict standardization of 103 sub-datasets. A pathological image processing strategy is proposed to assure homogeneity across images at the same scale, based on unified MPP.
- Inclusion of multi-scale images in the CPIA dataset, addressing the varying information present at different scales by pathologists. The CPIA dataset incorporates the diagnosis habits of pathologists for the first time to such an extent, resulting in a more clinically relevant dataset.
- Introduction of several independent pathology datasets and the CPIA-Mini dataset, enabling the researchers to explore the performance of various SSL pre-training methods.
- Evaluations with several SOTA SSL pre-training baselines and downstream fine-tuning, offering a solid foundation for future research in the field.

## II. CPIA Dataset

The CPIA dataset is a diverse and comprehensive collection of pathological images, serving as a valuable resource for deep learning researches. In this section, we will introduce the CPIA dataset from four aspects, including guiding principles, data assembly, data processing, and overview of the CPIA dataset.

All source datasets are either associated with the Creative Commons (CC) Licenses or encouraged sharing and derivatives, which allow us to develop the CPIA dataset based on them. The sub-datasets in the CPIA datasets use CC Licenses with the same License Elements as the source datasets.

### A. Guiding Principles

Existing studies have introduced a series of datasets, with each encompassing limited samples under diverse criteria. However, a dataset intended for SSL pre-training should ideally embody substantial data volume and comprehensive content coverage. Adhering to this concept, the CPIA dataset integrates a vast number of existing datasets and showcases several key characteristics, including diversity, extensiveness, standardization, multi-scale, and implementation flexibility. These characteristics will be introduced below.

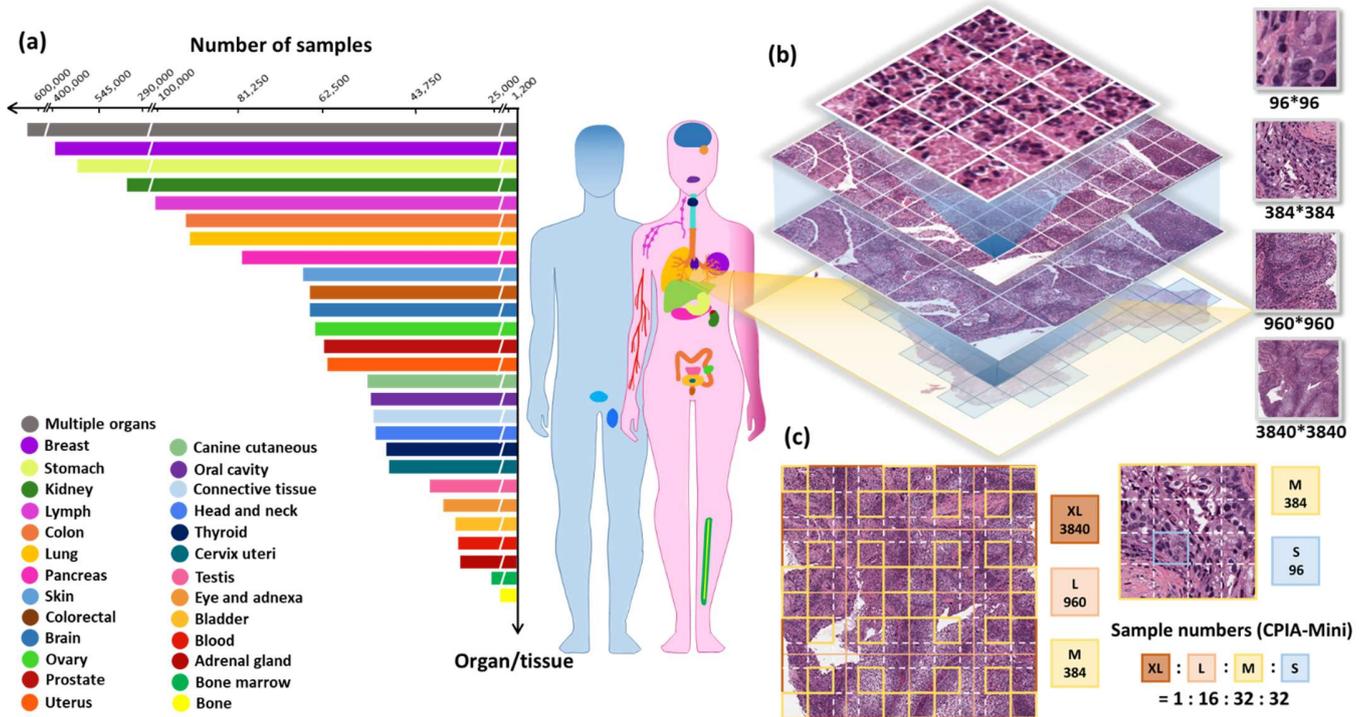

Fig. 1. The WSI processing strategy and compositions of the CPIA dataset. (a) Organ/tissue category composition and number of images in the CPIA dataset, which includes 21,427,877 standardized images, covering over 48 organs/tissues and approximately 100 kinds of diseases. The bar chart shows the composition of the CPIA-Mini dataset, and the full CPIA dataset is described in the "Overview of the CPIA Dataset" part in detail. (b) Sampling relationship between each scale of WSI sub-datasets. Specifically, 3840*3840 images represent the XL scale of WSI sub-datasets, which are directly cropped from the standardized WSIs, and the blank part is discarded. 960*960 and 384*384 images respectively represented the L and M scale of WSI sub-datasets, which are cropped from 3840*3840 images. The 96*96 images represent the S scale of WSI sub-datasets and are cropped from the 384*384 images. (c) The proposed sampling approach for the lightweight CPIA-Mini dataset, explains the 1:16:32:32 proportional relationship among XL, L, M, and S scales.

*1) Diversity*

The diversity in a pre-training dataset can enhance the generalization abilities of models. Therefore, to improve the performance across a wide range of individual downstream tasks, a comprehensive SSL pre-training dataset is needed [29].

The CPIA dataset shows a wide diversity demonstrated in various aspects. Specifically, the CPIA dataset comprises 103 sub-datasets obtained from distinct acquisition equipment, showcasing a variety of organs/tissues, diseases, and staining styles. **Fig. 1(a)** illustrates the composition of the CPIA dataset, which includes over 48 distinct organs/tissues, such as the breast, stomach, kidney, lymph nodes, colon, lung, and prostate. Furthermore, the CPIA dataset encompasses various disease types, covering approximately 100 categories, including breast carcinoma, colon adenocarcinoma, lung squamous cell carcinoma, prostate carcinosarcoma, clear cell renal cell carcinoma, etc. Lastly, the diversity of staining manifests in various ways. One is the heterogeneity arising from different staining conditions across various medical centers, leading to distinct hues even under the same staining method. The other heterogeneity is the utilization of numerous staining methods such as Hematoxylin-Eosin (H&E), Giemsa, and Immunohistochemistry (IHC), etc. To the best of our knowledge, the diversity of the CPIA dataset far exceeds other datasets in the literatures.

*2) Extensiveness*

The effectiveness of deep learning models is heavily influenced by the quantity and quality of training data [30]. The CPIA dataset not only has the diverse characteristic but also represents a significant advancement in large-scale pathological datasets. Specifically, the CPIA dataset contains 21,427,877 images, significantly surpassing all previous pathological datasets with magnitudes. The unprecedented scale of the CPIA dataset can not only support the training of next-generation large-scale models with comprehensive pathological knowledge but also offer a more valuable resource for future research efforts in discovering new patterns.

*3) Standardization*

The processing methods and organizational procedures of the CPIA dataset have been standardized. Currently, samples from various datasets may represent different objective scales, necessitating a significant workload to align them. To address this issue, two pathological image processing strategies have been established based on two data categories. Specifically, the processing method of WSI sub-datasets aligns samples based on MPP. Meanwhile, the ROI sub-datasets are selected artificially to align the WSI sub-dataset in different scales under the supervision of experienced senior pathologists. Therefore, different sub-datasets at the same objective scale reflect the same scale characteristics. Additionally, each sub-dataset is in accordance with the established standards for image segmentation or classification.

*4) Multi-scale*

As a diverse and extensive pathological dataset, the CPIA dataset necessitates a well-structured approach to be efficiently utilized. Unlike many existing datasets in literatures that only provide images at a single, predetermined scale tailored to a specific task, the CPIA dataset employs a unique approach.

The CPIA dataset is designed to reflect the clinical diagnosis habits of pathologists. They typically analyze pathological images at four different objective lens magnifications, each representing distinct pathological information. This dataset emphasizes the importance of multi-scale features in pathological images, which are often overlooked in existing datasets. In the CPIA dataset, image magnification scales are selected based on such clinical diagnosis habits of pathologists. Thus, this approach produces a more representative and clinically relevant dataset for SSL pre-training deep learning models in pathological image analysis [31]. And for the first time, the dataset construction aligns with the diagnosis habits of pathologists.

Consequently, sub-datasets at different scales each represent a unique diagnostic perspective, satisfying the need for multi-scale characteristics. By cooperating with experienced senior pathologists at Peking Union Medical College Hospital, we have developed a multi-scale strategy consistent with clinical practice. Thus, the CPIA possesses enhanced robustness and generalizability across numerous downstream tasks.

*5) Implementation Flexibility*

The CPIA is structured to the standard Image-Folder format, with each folder corresponding to a specific sub-dataset. This architecture simplifies implementation and allows researchers to conveniently filter or split the datasets based on particular downstream tasks.

Furthermore, the CPIA dataset is available in two versions: the full CPIA dataset and the CPIA-Mini dataset. The full CPIA dataset encompasses all processed data from the original datasets without any data filtering. The CPIA-Mini dataset serves as a lightweight version containing selected portions of each WSI sub-dataset, with samples resized to a uniform image size to reduce overall storage requirements. Serving as a lightweight version, the CPIA-Mini dataset contains 3,383,970 images. Meanwhile, to evaluate the SSL baselines on the proposed CPIA-Mini dataset, a series of additional independent pathological image datasets have been prepared. The validation and test subsets of these datasets are non-overlapping to the CPIA-Mini dataset. It is efficient and convenient to fairly evaluate the performance of various models on the proposed CPIA-Mini dataset. As a result, cost-effective approaches can be explored, offering valuable insights into the underlying pathological information and enabling fair comparisons between different methods. It also allows researchers to explore preliminary hyper-parameters on the CPIA-Mini dataset before launching SSL pre-training with the full CPIA dataset.

*B. Data Assembly*

The CPIA dataset has been curated to meet extensive and diverse requirements by gathering all publicly accessible pathological image datasets from existing literature. These datasets were initially disorganized and fragmented, necessitating an efficient sorting and stringent selection process to ensure the incorporation of only high-quality datasets. Details such as basic information and download links about these initial datasets are publicly available at

https://github.com/zhanglab2021/CPIA_Dataset, benefitting researchers in their respective fields.

After the data collection and cleaning stages, the CPIA dataset primarily consists of two major categories of pathological samples: WSIs and ROIs.

*1) WSIs*

WSIs are comprehensive pathological whole slide images captured by high-resolution scanners, which are widely applied in clinical analysis [32]. These digital images promote deep learning applications in pathological image analysis [4]. The CPIA dataset encompasses 55 WSI sub-datasets without strict annotations, covering over 45 organs/tissues and approximately 98 disease types. Most of the WSIs are tissue samples, collected to represent the histopathology sections of pathological image analysis.

*2) ROIs*

ROIs are representative images obtained under microscopes or specifically cropped from WSIs, being aligned after cleaning, standardization, and task-specific clustering [33]. The purpose of designing these ROIs is to accurately depict specific regions, thereby facilitating target pathological feature analysis in research. The CPIA dataset includes 50 ROI sub-datasets, covering over 29 organs/tissues and about 20 disease types. Most of the liquid samples are collected in the ROI group, representing the cytopathology sections of pathological samples.

### C. Data Processing

In order to accord with the diagnosis habits of pathologists, an effective multi-step strategy has been established to organize and process pathological images. This strategy aims to provide a more representative dataset with clinical workflows for pre-training, which makes up for the multi-scale features of pathological images ignored by existing datasets.

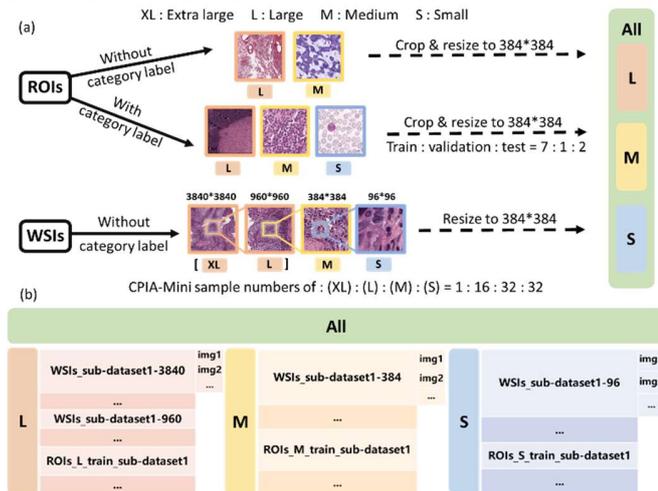

Fig. 2. (a) The CPIA-Mini dataset construction framework. (b) The internal framework of the data in the full CPIA and CPIA-Mini datasets. The L represents the L scale part of the CPIA dataset, named CPIA-L, including L scale ROI sub-datasets, and XL and L scale WSI sub-datasets. The M represents the M scale part of the CPIA dataset, named CPIA-M, including M scale ROI and WSI sub-datasets. The S represents the S scale part of the CPIA dataset, named CPIA-S, including S scale ROI and WSI sub-datasets.

In typical clinical practice, pathologists first start a diagnosis by scanning an entire sample section at the first level scale using a 4x objective lens, identifying tumors or suspicious lesions based on observed features. Secondly, the section is enlarged to the second level scale at a 10x objective lens to observe tissue morphology and marginal features. Then, at the third level scale, cell cluster characteristics and intercellular relationships are observed under a 20x objective lens. Lastly, at the fourth level scale, diseased cells are zoomed in to observe cellular characteristics with a 40x objective lens. This four-step process is integrated to complete the diagnosis of each sample slice, which is followed in the CPIA processing.

The CPIA dataset is designed to combine the diagnosis habits of pathologists with the data processing methods. There are two major categories of pathological samples collected in the CPIA dataset: WSIs and ROIs. Accordingly, we established two processing methods based on these two types of pathological images, with the supervision of experienced senior pathologists.

*1) WSIs*

To preserve information during data standardization, WSIs are resized using the MPP information attached to each sample. Due to the diversity of microscopes and high-resolution scanners, the original WSI sub-datasets have different MPP values. To standardize the publicly diverse WSIs, a unified MPP of 0.491 is determined to resize all WSIs, as it is widely used in data from the National Cancer Institute's Clinical Proteomic Tumor Analysis Consortium. Hence, all of the WSIs finally achieve the same physical scale per pixel.

Furthermore, inspired by the four-step diagnosis habits of pathologists, the WSIs are segmented into four scales: XL, L, M, and S (i.e., Extra-large, Large, Medium, and Small). With the supervision of experienced senior pathologists, the four selected scales can reflect key diagnosis scales of pathologists in clinical practice. Therefore, a four-step approach has been designed to achieve the cropping requirement.

Two sampling procedures are designed for the CPIA and CPIA-Mini datasets. To be specific, **Fig. 1(b)** displays the details of data sampling of the CPIA-Mini dataset, with XL scale samples cropped from resized WSIs under the resolution of 3840*3840. Subsequently, under the confirmation of pathologists, each XL sample is cropped into 16 images in L scale with a side length of 960, and 32 images in M scale with a side length of 384. Each image in M scale is then cropped an image in S scale with a side length of 96. To be mentioned, the white spaces of WSIs are discarded. Thus, the CPIA-Mini sample number ratio of XL, L, M, and S is about 1:16:32:32. **Fig. 1(c)** illustrates the selected cropping positions at different scales in each step. As shown in the "WSIs" section of **Fig. 2(a)**, in order to further lighten the CPIA-Mini dataset, the images of four scales are resized to 384*384. For the full CPIA dataset, all cropped sections are taken into account, except the S scale. Considering the balance between representativeness and enormity of the data, images in S scale are uniformly sampled 10% of them. Therefore, the full CPIA sample number ratio of XL, L, M, and S is about 1:16:100:160. The full CPIA dataset is not resized, making researchers use it flexibly according to their downstream tasks.

Each scale follows the scale hierarchy corresponding to the diagnosis habits of pathologists, allowing researchers to train models at multiple scales to obtain general pathological knowledge. Specifically, the XL scale enables the understanding of suspicious lesions, the L scale focuses on organizational characteristics composed of many cells, the M scale examines cell clusters and intercellular relationships, and the S scale explores individual cell morphological characteristics.

*2) ROIs*

Firstly, the ROIs are central cropped into the largest inscribed squares of the original images and then resized into 384*384. Then, due to the lack of a universally consistent MPP parameter, ROI sub-datasets are artificially divided into three scales: L, M, and S. The scale of each ROI sub-dataset is determined under the supervision of experienced senior pathologists. **Fig. 2** illustrates a structured approach designed to align ROIs to WSIs while maintaining similar feature characteristics. The L scale corresponds to the XL and L scales of the WSI sub-datasets. The M and S scales correspond to the same M and S scales of the WSI sub-datasets.

All the ROIs are kept in the full CPIA dataset, intended to meet the diverse needs of researches. Additionally, we have designed different ROI inclusion criteria to build the CPIA-Mini dataset. As shown in the "ROIs" section of **Fig. 2(a)**, only the training sets of the ROI sub-datasets with category labels are involved in the CPIA-Mini dataset. Furthermore, several independent downstream datasets are prepared, the validation and test subsets of which are non-overlapping to the CPIA-Mini dataset. Therefore, this design can support the efficiency and convenience of model evaluations.

In summary, the data processing strategy always adheres to the multi-scale characteristic. The internal framework of the CPIA and CPIA-Mini is the same. As shown in **Fig. 2(b)**, in order to merge the WSI and ROI sub-datasets, the framework of the CPIA dataset includes three levels: CPIA-L, CPIA-M, and CPIA-S, in three folders: L, M, and S. The L folder contains the XL and L scale WSI sub-datasets, as well as the L scale ROI sub-datasets. The M folder contains the M scale WSI sub-datasets and the M scale ROI sub-datasets. The S folder contains the S scale WSI sub-datasets and the S scale ROI sub-datasets. Each sub-dataset is in a single folder under L, M, and S folders for easy access. The specific sub-datasets and related information contained in each folder are available at https://github.com/zhanglab2021/CPIA_Dataset.

TABLE I
STATISTICS OF THE CPIA DATASET

| Row | Organ/tissue | Sub-dataset number | CPIA-L | | | CPIA-M | | CPIA-S | |
|---|---|---|---|---|---|---|---|---|---|
| | | | WSIs-XL | WSIs-L | ROIs-L | WSIs-M | ROIs-M | WSIs-S | ROIs-S |
| 1 | Adrenal gland | 1 | 358 | 5,559 | 0 | 10,961 | 0 | 10,740 | 0 |
| 2 | Bladder | 1 | 397 | 6,046 | 0 | 11,823 | 0 | 11,507 | 0 |
| 3 | Blood | 8 | 0 | 0 | 0 | 0 | 0 | 0 | 51,604 |
| 4 | Bone | 1 | 0 | 0 | 1,144 | 0 | 0 | 0 | 0 |
| 5 | Bone marrow | 2 | 902 | 3,665 | 0 | 23,200 | 0 | 18,787 | 1,004 |
| 6 | Brian | 2 | 1,596 | 20,760 | 0 | 40,944 | 0 | 39,276 | 0 |
| 7 | Breast | 13 | 3,884 | 50,073 | 279,861 | 207,970 | 10,105 | 191,749 | 0 |
| 8 | Canine cutaneous | 1 | 1,000 | 15,185 | 0 | 29,818 | 0 | 28,791 | 0 |
| 9 | Cervix uteri | 1 | 844 | 12,739 | 0 | 24,783 | 0 | 24,111 | 0 |
| 10 | Colon | 5 | 5,507 | 46,427 | 0 | 257,978 | 16,398 | 219,559 | 0 |
| 11 | Colorectum | 10 | 115 | 1,783 | 3,540 | 3,525 | 142,939 | 3,464 | 0 |
| 12 | Connective tissue | 1 | 4,034 | 48,856 | 0 | 302,175 | 0 | 288,907 | 0 |
| 13 | Eye and adnexa | 1 | 474 | 7,187 | 0 | 14,131 | 0 | 13,941 | 0 |
| 14 | Head and neck | 1 | 6,036 | 73,071 | 0 | 450,970 | 0 | 432,351 | 0 |
| 15 | Kidney | 5 | 15,683 | 212,967 | 0 | 1,126,696 | 1,419 | 1,079,566 | 0 |
| 16 | Liver | 1 | 5,696 | 83,922 | 0 | 520,918 | 0 | 509,939 | 0 |
| 17 | Lung | 5 | 51,790 | 683,060 | 0 | 4,231,568 | 25,171 | 4,026,259 | 0 |
| 18 | Lymph | 4 | 3,492 | 37,928 | 374 | 214,733 | 220,025 | 203,841 | 0 |
| 19 | Oral cavity | 1 | 952 | 14,748 | 0 | 28,965 | 0 | 28,277 | 0 |
| 20 | Ovary | 2 | 3,031 | 27,486 | 0 | 109,387 | 0 | 95,423 | 0 |
| 21 | Pancreas | 2 | 5,536 | 69,566 | 0 | 372,647 | 0 | 357,403 | 0 |
| 22 | Prostate | 3 | 1,000 | 15,592 | 331 | 30,440 | 18,783 | 29,168 | 0 |
| 23 | Skin | 3 | 5,898 | 68,467 | 0 | 396,103 | 0 | 377,725 | 0 |
| 24 | Stomach | 4 | 876 | 13,577 | 1,770 | 26,691 | 224,364 | 25,997 | 146,651 |
| 25 | Testis | 1 | 869 | 8,842 | 0 | 17,372 | 0 | 15,597 | 0 |
| 26 | Thyroid | 1 | 855 | 13,265 | 0 | 25,964 | 0 | 25,290 | 0 |
| 27 | Uterus | 3 | 9,024 | 108,317 | 0 | 650,964 | 0 | 611,392 | 0 |
| 28 | Multiple organs | 20 | 10,137 | 139,978 | 521 | 275,661 | 12,992 | 264,382 | 0 |
| 29 | Total number | 103 | 139,986 | 1,789,066 | 287,541 | 9,406,387 | 672,196 | 8,933,442 | 199,259 |
| | Total number of the CPIA dataset: 21,427,877 | | | | | | | | |

*D. Overview of the CPIA Dataset*

*1) Sample Statistics*

The CPIA dataset contains 21,427,877 images that are divided into 28 categories by organs/tissues, as shown in **Table I**. Row 1-27 of **Table I** are specific organs/tissues, such as adrenal gland, bladder, blood, bone, etc. Each organ/tissue category covers multiple sub-datasets and diverse related diseases. Row 28, the multiple organs category, contains more than one organ/tissue in its sub-datasets.

*2) Thumbnail of the CPIA Dataset*

**Fig. 3** offers a thumbnail summary of the CPIA dataset, highlighting its diversity, standardization, and multi-scale characteristics. Regarding the diversity characteristic, the randomly selected sub-datasets include different cells and tissues reflecting organ and disease diversity. And the different hues and staining methods intuitively reflect staining style diversity. Regarding the standardization characteristic, different sub-dataset images in the same column have the same scale characteristics. Regarding the multi-scale characteristic, from left to right, the scale of the pathological images is gradually enlarged. These multiple scales respectively represent the key diagnostic perspectives of pathologists.

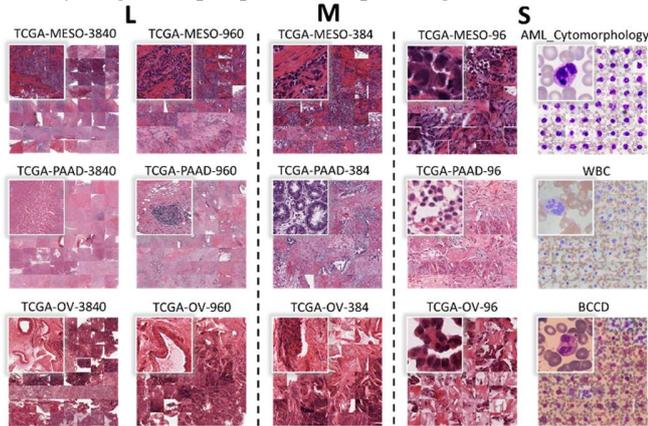

Fig. 3. A thumbnail summary of the CPIA dataset, showing the diversity, standardization, and multi-scale characteristics. From left to right, the scale of the pathological images is gradually enlarged. Where the first and second columns represent the image features in the L folder, the third column represents the image features in the M folder, and the fourth and fifth columns represent the image features in the S folder.

*3) Naming Rule of Processed Images*

In the data processing flow of the WSI sub-datasets, this paper has established a naming rule for the processed images to link them with the original images and to show the cropping relationship between them. Each processed image is named in the format of "original name-cropped size-X axis displacement unit-Y axis displacement unit". For example, an image named "TCGA-A2-A0EY-01A-01-BSA-960-39-6" is cropped from the original image named "TCGA-A2-A0EY-01A-01-BSA", and the top left corner of the cropped image is located at (960*39, 960*6) in the original WSI.

In the data processing flow of the ROI sub-datasets, each ROI sub-dataset is aligned to a specific scale, and the image names of the ROI sub-datasets are consistent with the original images.

*4) The CPIA-Mini Dataset*

The CPIA-Mini dataset is used in the experiment. It consists of a select fraction from each WSI sub-dataset, all samples from ROI sub-datasets without category labels, and training sets of ROI sub-datasets with category labels. **Fig. 2(a)** illustrates the workflow and framework of the CPIA-Mini dataset, which closely resembles the full version but involves resizing operations of WSIs.

Further details of the CPIA-Mini dataset are not elaborated in this section. For additional information, please refer to https://github.com/zhanglab2021/CPIA_Dataset.

## III. EXPERIMENTS

In order to take a preliminary evaluation of the CPIA dataset and explore the reasonability of its multi-scale framework, this section conducts a series of experiments. We first conducted SSL pre-training with the CPIA dataset and then fine-tuned pre-trained models on the downstream tasks. To be specific, SSL pre-training stage includes two main paradigms: contrastive learning and reconstructive learning. In the fine-tuning stage, there are three downstream datasets representing the scales of main downstream tasks in the field of pathological image analysis. After the SSL pre-training based on the CPIA dataset, the performance of models in multiple downstream tasks proves the effectiveness of the CPIA dataset and the reasonability of its multi-scale framework. For the reason that the CPIA dataset is gradually updated and refined, the experiments were based on the CPIA-Mini dataset, which is referred to as the CPIA dataset in the experiments section.

*A. Experimental Design*

Two types of SSL frameworks have demonstrated advanced performance in the field of computer vision. The first type is contrastive learning, in which models primarily learn useful representations by contrasting between positive and negative instances [34-41]. The second type is reconstructive learning, in which models learn general features and representations from the image by focusing on image reconstruction tasks [42]. These two types have been applied to three widely utilized pre-training methods, which are deployed as our baselines:

MoCo-v3 is a contrastive learning framework using Vision Transformer (ViT) as its backbone [41]. This framework utilizes a prediction head and eliminates the memory queue mechanism [43]. During the training process, MoCo-v3 employs a fixed random patch projection layer, which enhances the training stability and leads to improved training outcomes.

DINO also employs the popular ViT as its backbone [44]. However, unlike traditional contrastive learning frameworks that take negative samples, DINO leverages a knowledge distillation strategy. This method employs a teacher network to generate augmented versions of the training data, then trains the student model to learn from the teacher network.

MAE method uses a reconstructive learning framework based on masking [42]. This method first split an image into patches for random masking under a certain ratio. A ViT encoder is then employed to encode the unmasked patches, followed by a lightweight decoder restoring the masked positions.

To investigate the performance enhancement of the CPIA dataset under the SSL process, this work employs three experiments to compare: (i) randomly initialized ViT models without pre-training; (ii) randomly initialized ViT models using the ImageNet-1k dataset under the MAE method; and (iii) randomly initialized ViT models using the CPIA dataset under the MAE method.

Further experiments are also designed to explore the impact of various model initializations and SSL methods: (i) random initialization, followed by pre-training using the MAE, MOCO-v3, and DINO methods; (ii) TIMM official pre-trained with ImageNet-1k, followed by pre-training using the aforementioned three methods [45].

Additionally, the performances of sub-datasets at different scales in the CPIA dataset are compared: models pre-trained with ImageNet-1k are used for SSL with CPIA-L, CPIA-M, and CPIA-S datasets using the MAE method.

Finally, all the processed ViT models have their parameters extracted and loaded into the ViT backbone model built with TIMM, which are subjected to fine-tuning experiments for three distinct downstream tasks.

### B. Fine-tuning Datasets

Following the SSL process, the models need to be fine-tuned to assess their performance in downstream classification tasks. The classification results preliminary evaluate the data modeling capability. Furthermore, three independent datasets representing various scales are specially employed, which offer diverse perspectives regarding the multi-scale design of CPIA.

The Raabin-WBC dataset is abbreviated as WBC in this paper [46]. This S scale dataset is composed of microscopic images of white blood cells, with each image containing only one or two stained white blood cells. For the downstream classification tasks, this dataset comprises 301 basophil, 1066 eosinophil, 3461 lymphocyte, 795 monocyte, and 8891 neutrophil images.

The pRCC dataset consists of Papillary Renal Cell Carcinoma subtyping images, selected and cropped by pathologists from the TCGA-KIRP dataset [33]. This dataset comprises 870 type 1 ROIs and 547 type 2 ROIs, with each image meeting the M scale dataset criteria.

The Camelyon16 dataset is abbreviated as CAM16 in this paper. This WSI dataset is derived from the Cancer Metastases in Lymph Nodes challenge [23]. In each WSI, we select 5 to 10 ROIs with dimensions of 8000*8000 (2560*2560 under the CPIA standard) to meet the average L scale dataset criteria. Our CAM16 dataset comprises 540 tumor and 541 normal images.

In the pRCC and CAM16 datasets, each category is partitioned into training, validation, and test sets with a ratio of 7:1:2. Notably, the official website of the WBC dataset provides two test sets, and we select Test-A as the designated test set. Additionally, we allocate a segment of the official training set as a validation set, resulting in a final ratio of 10:2:5 for the training, validation, and test sets respectively for each cell type.

### C. Experimental Implementation

The pre-training experiment is conducted by implementing three experimental models in Python 3.8 with official codes. The versions of the deep learning frameworks are selected to meet the official requirement of each method. MAE and MOCO-v3 use PyTorch 1.9.0, Torchvision 0.10.0, and CUDA 10.2. DINO uses PyTorch 1.7.1, Torchvision 0.8.2, and CUDA 11.0. Each model is trained on a server with two Nvidia A100-SXM4-80GB GPUs. The batch size per GPU is set to 1024 for MAE, 512 for MOCO-v3, and 256 for DINO.

For the downstream task experiments, we use a server with one Nvidia RTX 3090 GPU, and the model backbone is implemented in Python 3.8 with PyTorch 1.10.0, Torchvision 0.10.0, and CUDA 11.3. Each experiment undergoes 50 training epochs, during which the selection of the best-performing epoch on the validation data determines the output model for the downstream task. The final results of the experiments are derived from the classification outcomes of the downstream tasks on the independent test sets. Each image that goes through the pre-training or fine-tuning process has the input size been set to 3*224*224.

TABLE II
EXPERIMENTAL RESULTS

| Model | | | WBC | | pRCC | | CAM16 | |
|---|---|---|---|---|---|---|---|---|
| Initial Setting | Method | Dataset | Acc (%) | F1 (%) | Acc (%) | F1 (%) | Acc (%) | F1 (%) |
| Random | None | None | 91.63 | 85.35 | 73.85 | 71.12 | 80.09 | 79.94 |
| Random | MAE | ImageNet-1k | 97.17 | 95.36 | 91.17 | 90.66 | 93.98 | 93.98 |
| Random | MAE | CPIA | **97.58** | **95.83** | **92.58** | **91.94** | 93.98 | 93.98 |
| Random | MOCO-v3 | CPIA | 91.03 | 85.70 | 73.14 | 68.05 | 81.02 | 80.76 |
| Random | DINO | CPIA | 91.24 | 85.31 | 70.32 | 62.83 | 80.09 | 79.94 |
| ImageNet-1k | MAE | CPIA | 97.35 | 95.66 | 92.23 | 91.82 | 91.67 | 91.65 |
| ImageNet-1k | MOCO-v3 | CPIA | 90.92 | 86.72 | 75.62 | 73.38 | 80.56 | 80.31 |
| ImageNet-1k | DINO | CPIA | 89.65 | 85.38 | 71.73 | 66.81 | 80.09 | 79.94 |
| ImageNet-1k | MAE | CPIA-L | 97.33 | 95.73 | 89.40 | 88.60 | **92.59** | 92.58 |
| ImageNet-1k | MAE | CPIA-M | 97.14 | 95.12 | **89.75** | 89.04 | 92.59 | 92.59 |
| ImageNet-1k | MAE | CPIA-S | **97.49** | 95.57 | 84.45 | 83.28 | 92.13 | 92.13 |

## D. Experimental Results

*1) The Performance of the CPIA Dataset*

The top 3 rows of **Table II** present the classification results of three distinct models performing on three downstream datasets. The initial three rows characterize the performance of each model: the first model serves as a baseline with completely random parameters. The second model, initialized randomly, is officially pre-trained on MAE with the ImageNet-1k dataset. Conversely, the third randomly initialized model utilizes the CPIA dataset as the pre-training dataset.

The control group, a randomly initialized model without pre-training, only achieves accuracies of 91.63%, 73.85%, and 80.09% on WBC, pRCC, and CAM16. In comparison, the model pre-trained with the ImageNet-1k dataset achieves accuracies of 97.17%, 91.17%, and 93.98% on the three tasks. And the model pre-trained with the CPIA dataset outperforms the former one by achieving accuracies of 97.58%, 92.58%, and 93.98%. The different results between the pre-trained models and the control group indicate that both pre-training processes provide improvements in model performance.

Despite undergoing more comprehensive official training processes, the general vision-based model still demonstrates relatively poor performance in downstream tasks compared with the models pre-trained with the CPIA dataset. This is attributed to the significant domain gap between the pre-training and fine-tuning phases. The CPIA dataset enables the model to better learn pathological features during pre-training, which boosts domain-specific performance in downstream tasks. These insights underscore the imperative role of the CPIA dataset in pathological image analysis.

*2) Observations on Domain Gap between Different Initializations*

Introducing general vision knowledge before the CPIA dataset training may improve the overall comprehension of the model. Therefore, the CPIA can be used to bridge the general pre-training and domain-specific learning. To explore the effectiveness of pathological knowledge encoding with the CPIA dataset, we implemented the experiments with random initialization and natural image pre-training [45]. The models are first initialized before applying SSL training methods with CPIA. Illustrated in the middle part of **Table II**, the ViTs pre-trained on MAE obtained better downstream performance than MOCO-v3 and DINO over the three tasks (a higher F1 score at 10% in general). Focusing on the SOTA MAE method, the randomly initialized model outperforms the ImageNet initialized one by 0.23%, 0.35%, and 2.31% in accuracies. The results show that a significant domain gap between natural and pathological images leads to conflicts in prior knowledge. This, in turn, can diminish downstream performance. In pathological image analysis, pre-training research is urged to better merge general vision knowledge with domain-specific expertise. As a large-scale foundational dataset, the CPIA dataset facilitates such exploratory endeavors in future studies.

*3) Observations on CPIA Feature Scales*

This section further explores the effect of multi-scale subdivisions within the CPIA dataset. To be mentioned, the SOTA MAE method is applied before task-specific fine-tuning. The lower part of **Table II** shows the performance of models pre-trained with different CPIA subsets. The model achieves accuracies of 97.49%, 89.75%, and 92.59% in downstream WBC, pRCC, and CAM16 tasks when pre-trained with the corresponding S, M, and L scales. These results surpass the performance of models pre-trained on non-matching scales. Moreover, for the WBC, the model pre-trained with the corresponding scale subset outperforms the model pre-trained with CPIA by 0.14% in accuracy. Similarly, for CAM16, the performance improvement amounts to 0.92%.

These outcomes highlight the performance disparity across different CPIA subsets, underscoring the efficacy of the CPIA multi-scale standards. In the domain of pathological image analysis, due to the presence of multi-scale tasks, a more advanced pre-training method is necessary. It should effectively balance feature variety and inductive bias across multiple scales. The CPIA dataset provides an efficient scale subdivision standard that enables scale-related pre-training explorations.

*4) Observations on SSL Methods*

This section evaluates the efficacy of various pre-training methods on CPIA. As demonstrated in the middle panel of **Table II**, models pre-trained by MAE significantly surpass MOCO-v3 and DINO in fine-tuning. Initialized with random weights and ImageNet-1k pre-training, MAE consistently bests MOCO-v3 and DINO by roughly 6.5%, 20%, and 10% in accuracies on WBC, pRCC, and CAM16. Additionally, MAE achieves these superior results at a faster iteration speed.

These results indicate that, under the present conditions, the MAE method demonstrates considerable advantages in both performance and training efficiency. In the case of contrastive learning based DINO and MOCO-v3, two parallel models participate in training, leading to significant time and memory consumption. Conversely, the masking learning strategy of MAE allows the model to process only a portion of the image data, significantly reducing data volume and making the training process lighter and more efficient. As we further investigate the pathological pre-training, the CPIA dataset may pave the way for fast and lightweight reconstruction methods for SSL in future work.

## IV. Conclusions

In this work, we present the CPIA dataset, the first highly diverse dataset for SSL pre-training on pathological images. The CPIA dataset is rigorously standardized, comprising 21,427,877 images that cover over 48 organs/tissues and approximately 100 kinds of diseases. With the supervision of experienced senior pathologists, the multi-scale standardization enables the models to acquire a more general and explainable understanding of pathological image analysis. Several SOTA baselines of SSL pre-training and various downstream evaluations were conducted using the CPIA dataset. The results indicate the potential usage of the CPIA dataset and highlight the diversity of information hidden under images at different scales. Our explorations may shed light on new researches focused on multi-scale related pre-training and pattern discovery explorations. We also believe that the CPIA dataset can provide a new extensive benchmark, thus laying a solid foundation for the pre-training studies of large-scale models in the field of pathological image analysis.